# Mesoporous Silica as a Carrier for Amorphous Solid Dispersion

## Smruti P. Chaudhari[1*] and Anshul Gupte[1]

*[1]Mayne Pharma Inc, Greenville, NC, USA.*

***Authors' contributions***

*This work was carried out in collaboration between both authors. Both authors read and approved the final manuscript.*

| Review Article |
| --- |

## ABSTRACT

In the past decade, the discovery of active pharmaceutical substances with high therapeutic value but poor aqueous solubility has increased, thus making it challenging to formulate these compounds as oral dosage forms. The bioavailability of these drugs can be increased by formulating these drugs as an amorphous drug delivery system. Use of porous media like mesoporous silica has been investigated as a potential means to increase the solubility of poorly soluble drugs and to stabilize the amorphous drug delivery system. These materials have nanosized capillaries and the large surface area which enable the materials to accommodate high drug loading and promote the controlled and fast release. Therefore, mesoporous silica has been used as a carrier in the solid dispersion to form an amorphous solid dispersion (ASD). Mesoporous silica is also being used as an adsorbent in a conventional solid dispersion, which has many useful aspects. This review focuses on the use of mesoporous silica in ASD as potential means to increase the dissolution rate and to provide or increase the stability of the ASD. First, an overview of mesoporous silica and the classification is discussed. Subsequently, methods of drug incorporation, the stability of dispersion and, much more are discussed.

*_________________________________________________________________________________________*

*\*Corresponding author: E-mail: smruti.chauhari@maynepharma.com;*

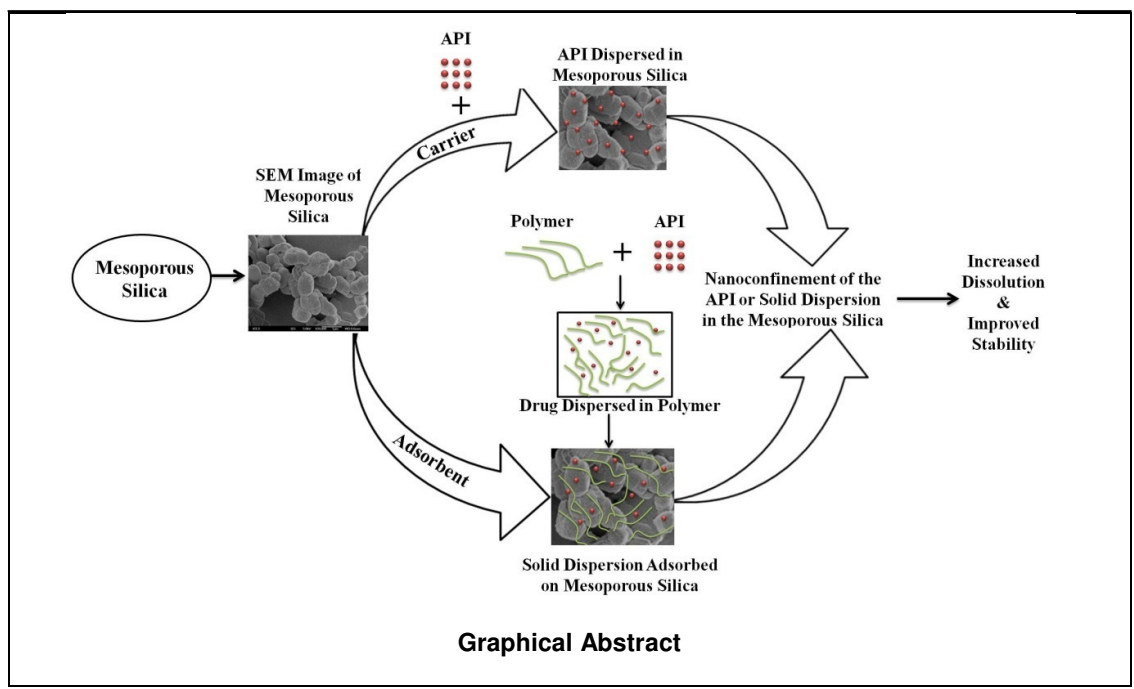

**Graphical Abstract**

## 1. INTRODUCTION

Combinatorial chemistry and high-throughput screening of drugs has led to the discovery of more and more poorly water soluble drugs [1-3]. These are target specific drugs with poor aqueous solubility, which present a challenge for effective drug delivery. Many strategies have been explored to overcome their low bioavailability because of poor solubility. Converting crystalline drugs into an amorphous form is one the most promising tool to overcome the problem of poor solubility, as compared to simple micronization of drugs [4] or salt formation [5], which has many practical limitations. The process of amorphization of the drug substance involves disruption of the ordered structure of a crystal which can be achieved by using a process like melt cast, hot melt technology and spray drying [6,7]. Hence, amorphous drug exists at a higher free energy state and has higher mobility as compared to their crystalline counterpart. The amorphous drug has enhanced thermodynamic properties which lead to higher apparent solubility and dissolution rate [8-10]. Due to high internal energy and enhanced molecular mobility of amorphous system, it has a tendency to crystallize during storage and dissolution [11-13]. In order to benefit from the amorphous drug and to reduce the conversion into crystalline form, efforts have been made to reduce the driving force of recrystallization by use of precipitation inhibitors [14,15]. The chemical potential of the drug in crystalline phase is equal to the chemical potential on the liquid phase at melting point ($T_m$) of the crystalline drug. However, the chemical potential of the amorphous drug is high, which can be lowered by the addition of polymer. The drug dispersed in an inert water soluble carrier like polymer at the solid stage is known as solid dispersion. It is used to improve solubility of amorphous solid dispersion. Hence, the concept of solid dispersion is gaining momentum. The concept of solid dispersion was introduced by Sekiguchi and Obi [16] and it has become the preferred and most successful method to enhance the drug dissolution and to stabilize the amorphous solid dispersion (ASD). To achieve stable ASD, a single phase system with drug molecules miscible with the polymer in a single phase forming a glass solution is desired [17-21]. The physical stability of ASD has been attributed to several factors. The addition of polymer to ASD will help to stabilize the system by reducing the mobility of the drug molecules and by increasing the glass transition temperature of the glass solutions as compared to pure amorphous drugs, and kinetically act as crystallization inhibitors [22-24]. Some scientists have also proved that the drug-polymer interactions are important for stabilization of ASD [22,25,26].

Another approach to stabilizing an amorphous drug delivery system is to use porous media or adsorbents. Porous media refers to materials with large surface areas (e.g. greater than 500 $m^2$/g) and pore volumes (e.g. greater than 1 $cm^3$/g) [18]. The International Union of Pure and Applied Chemistry (IUPAC) classified the porous media is based on pore size; micropore, mesopore, and macropore. Micropore has a mean pore diameter less than 2 nm, mesopores have a pore diameter in between 2- 50 nm and macropore have a pore diameter greater than 50 nm.

In the context of Noyes-Whitney equation [27], an enhancement of dissolution rate can be achieved by an increase in apparent solubility, reduced boundary-layer thickness and higher specific surface area [12,28]. Konno T and his co-workers demonstrated the use of porous media as a potential drug delivery system to improve solubility and dissolution rate [29-32]. The porous media may be advantageous in enhancing the physical stability of amorphous systems since it has a size constraint effect on the nucleation and crystal growth. According to the classical theory of homogeneous nucleation, crystal growth proceeds spontaneously once a critical nucleation size is reached. The system will exist in a non-crystalline state if nucleation and growth are prevented. This can be achieved if spatial constraints of a capillary are imposed [33]. Jackson and McKenna investigated the mechanism of vitrification using porous adsorbents with controlled pore glasses (CPG) with mean pore diameters ranging from 4 to 73 nm and crystalline o-terphenyl as a model system [34]. Crystallization of o-terphenyl is not seen in CPG with 4nm pores, regardless of whether the pores are overfilled or underfilled. Since the critical nucleation size is 6.2 nm which is greater than the pore diameter of 4 nm i.e. there were not enough molecules in the pores to reach the critical nucleus and nucleation could not proceed. Hence, nanoconfinement of drugs prevents recrystallization of the drugs thereby stabilizing the amorphous system. It was found that the drug cannot re-crystallize when the space in which it is confined does not exceed the drug molecule width by at least a factor of 10 [35, 36].This unique process can be utilized to deliver the poorly soluble drugs.

The behavior of drug molecules confined to porous media is different than those in bulk. In the case of drug molecules confined to porous media, the drug molecules have an altered thermodynamic state which plays an important role in the physical stability and product performance. It may also influence the chemical stability as well. The purpose of this review is to focus on the use of mesoporous silica in amorphous oral drug delivery system. Mesoporous silica can be classified as ordered mesoporous silica and non-ordered mesoporous silica.

## 2. ORDERED MESOPOROUS SILICA

Ordered Mesoporous silica (OMS) are porous materials that exhibit an array of uniform mesopores. These have a pore diameter between 2 and 50 nm. OMS exhibit a very high specific surface area (up to 1500 $m^2$/g) and pore volume (up to 1.5 $cm^3$/g) and hence they have high porosity. OMS (MCM- Mobil Composition of Matter) were first synthesized by Mobil Oil Corporation in 1991. Since then, ordered mesoporous silica is attracting attention for applications in fields such as lasers, sensors, catalysis and solar cells [37,38]. Due to the unique features of silica based ordered mesoporous materials, they are excellent candidates for controlled drug delivery systems [39]:

- These materials have ordered pore network, which is homogeneous in size and allow control of the drug loading and release kinetics for the drug
- It has a high pore volume which can fit in large amount of pharmaceuticals
- It has silanol containing surface that can be functionalized to allow better control over drug loading and release.

Vallet-Regi research team proposed the use of ordered mesoporous silica as a drug delivery system on model drug Ibuprofen [38]. The pore channel of MCM-41 was loaded with ibuprofen and this system release the drug in a sustained manner. Since then, there has been a tremendous increase in the use of the mesoporous system as a drug delivery system. A significant growth and many efforts have been devoted to tailoring the nanostructure and textural properties which allow control over drug loading and release kinetics. To confine the large size molecule such as protein, Vallet-Regi developed a method for tailoring pore sizes of Santa Barbara series (SBA-15) mesoporous material. They have used hydrothermal treatment to enlarge the pore diameters. The mesoporous material synthesized using this novel method were named SBA-15-3d, SBA-15-5d, and SBA-15-7d, the number 3d, 5d and 7d stand for the total time of hydrothermal treatment in the autoclave [40]. The increasing pore size increased the drug loading and protein adsorption.

Various studies have demonstrated enhanced *in vitro* dissolution properties of silica based materials for a wide variety of compounds. The mesoporous silicates improve the drug absorption by creating supersaturation *in vivo* [41]. In the context of Fick's first law [42], increased intraluminal concentration may enhance the flux of drug across the gastrointestinal membrane and which eventually increases absorption. However, for optimal *in vivo* performance simply releasing the drug in a supersaturated state may not be sufficient, since rapid precipitation of drugs to more stable crystalline form may jeopardize the absorption [43-47]. If we maintain the supersaturated state for a sufficient period, it will result in increased absorption which in turn increases bioavailability. The addition of excipients which prevent and retard precipitation will be beneficial. This type of approach, i.e. in which the formulation components generate a supersaturated state of drug (Spring) and formulation component which stabilize the supersaturated solution, precipitation inhibitors (parachute) is known as Spring and Parachute approach [43,48]. Various dissolution enhancing technologies that are associated with the creation of supersaturation have been shown to benefit from incorporation of precipitation inhibitors, for e.g. solid dispersion [49,50], lipid base systems [44-47] or use of high solubility salts [43]. Speybroeck et al has demonstrated that the approach of combining SBA-15 with an appropriate precipitation inhibitor is established as a valuable tool to enhance the absorption of poorly soluble weak base itraconazole (ITZ) [51]. The pharmaceutical performances of two precipitation inhibitors (hydroxypropyl methyl cellulose (HPMC) and hydroxypropyl methyl cellulose acetate succinate (HPMCAS)) were studied. It was found that the addition of HPMC polymer proved to be a good stabilizer for ITZ supersaturation in intestinal media and the addition of HPMC to SBA-15 resulted in more than 60% increase in absorption [51]. Ambrogi et al investigated the effect of ordered mesoporous silica like MCM-41 and SBA-15 [52,53] on carbamazepine. The dispersion prepared was amorphous in nature and showed improved dissolution and physical stability of the drug.

## 2.1 Preparation of Ordered Mesoporous Silica

The first ordered mesoporous silica synthesis is based on the use of a surfactant as structure directing agent for the silica [54,55]. At critical micelle concentration (CMC) in aqueous solution, surfactant molecules start to form aggregates, which are known as micelles. The shape and size of the micelle depend on several factors, such as chemical composition, auxiliary chemicals, nature of surfactant and reaction conditions such as concentration, pH and temperature. These micelles aggregate to form supramicellar structures which, depending on the conditions, can be cubic, hexagonal or lamellar to determine the final mesopores framework. The mechanism of formation of this material takes place by means of "liquid-crystal templating" (or cooperative self-assembly), in which the silicate material forms inorganic walls between ordered surfactant micelles, forming a template framework. The surfactant is removed by solvent extraction or calcination process at high temperature, this is the last stage of the process. As a consequence of template synthesis, a material with narrow pore size distribution and ordered distribution of mesopores are achieved. A range of new materials with an ordered distribution of mesopores, narrow pore size distribution, homogeneous pore morphology (hexagonal and cubic pores), high surface area ($1000m^2/g$), tuneable pore size (2-10 nm) and high pore volume (1 $cm^3/g$) has been synthesized. The pore diameter can be varied from 2 to 10 nm by the addition of auxiliary hydrocarbons such as alkylated benzene or by changing the alkyl chain length [40]. There are several types of mesoporous materials discovered, depending on the synthesis procedure for e.g. SBA series [56-58], MCM series [55,59-63], KIT-1 (Korea advanced institute of Science and Technology no. 1) [64], MSU series (Michigan State University) [65],and FSM-16 (Folded Sheet Material no.16) [66,67]. Fig. 1 shows the TEM image of most popular SBA-15 [68].

## 2.2 Stability of Drug Loaded on Ordered Mesoporous Silica

The stability of ITZ formulation based on ordered mesoporous silica SBA-15 has been studied by

R. Mellaerts, et al. [69]. Upon contact with the simulated gastric fluid, ITZ is readily liberated from the pore space of SBA-15 carrier material; a supersaturated ITZ solution is obtained. When this dispersion was stored at 0% RH, the performance of the ITZ is maintained. Interestingly, at higher humidity condition of 52% and 97% RH, an even higher supersaturation level is obtained when the aged formulation is suspended in simulated gastric fluid. This is a very intriguing feature of this formulation. Conversely, solid dispersion made using polymer like PVP and HPMC, a decrease release performance on storage is observed due to crystallization of the drug on storage. Mellaerts R demonstrated that under humid conditions, SBA-15 undergoes hydroxylation, which concentrates the SBA-15 surface more hydrophilic, which facilitates the liberation of poorly water soluble ITZ molecules upon competitive adsorption of water molecules in simulated gastric fluid as shown in Fig. 2 [69]. Thus, it is possible to tune the release profile of poorly water soluble drugs by varying SBA-15 hydrophilicity. It is hypothesized that at high drug loading of ITZ, the walls of SBA-15 are coated with drug molecules and thus there are structural changes that occur upon its storage at the humid condition, specifically the void space of SBA-15 are suppressed and thus the aggregation of hydrophobic ITZ is prevented [69]. Limnell et al [70] studied the stability of high drug load indomethacin formulations based on ordered mesoporous silica, MCM-41, and SBA-15 materials. After drug loading, only a small amount of crystalline materials (<3.0 wt.-%) were detected by DSC. The physicochemical stability of the high drug load indomethacin formulation was found to be variable during prolonged storage under stressed conditions. Even though the physical stability of the samples was found satisfactory and drug release performance was rapid after storage. The dissolution studies revealed the ability of both MCM-41 and SBA-15 microparticle to improve the dissolution of indomethacin by stabilizing it in a disordered state inside the mesopores and by further maintaining this state even during storage in accelerated conditions [71]. The chemical analyses show a decrease in the loading degree and possible indomethacin degradation product formation, especially in MCM-41 based samples after stressing. The indomethacin SBA-15 formulation is more stable under stress condition, this result is in agreement with the study conducted by Van Speybroeck, et al. [72]. Ten (10) physicochemically diverse model compounds were successfully loaded onto SBA-15. In all cases, the adsorbed drug fraction was found to be amorphous, as evidenced by DSC results. Amorphization of the compounds resulted in a significant improvement in the dissolution rate as compared to the crystalline drugs. On the storage of formulation at 25°C and 52% RH, no drug crystallization was observed after 6 months and the pharmaceutical performance of the SBA-15 formulation was retained accordingly. The authors concluded that this established the potential of SBA-15 to yield physically stable, dissolution enhancing formulation, irrespective of a drug's physicochemical profile.

# 3. NON-ORDERED MESOPOROUS SILICA

Various non-ordered mesoporous silica like Syloid, Sylysia and aeroperl have been used to improve the dissolution properties of poorly soluble drugs. This mesoporous silica has advantages over non-porous high surface area materials, in which the deposited molecules not just are adsorbed on the surface, but also confined to pore that are only a few molecular diameters wide hence preventing recrystallization [73-75]. Takeguchi and coworkers have evaluated the solid dispersion prepared using porous silica-sylysia 350 like nonporous Aerosil 200 on model drug indomethacin [76]. The dissolution rate of indomethacin was faster in the solid dispersion prepared using sylysia as compared to solid dispersion prepared using aerosil. A similar study was conducted with model drug tolbutamide [77], bicalutamide [78], resveratrol [79] and the results were in agreement with the previous study conducted.

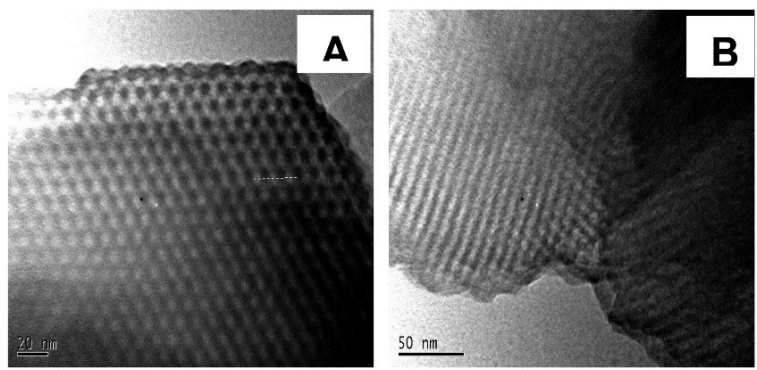

**Fig. 1. TEM images of mesoporous silica SBA-15; Reprinted from reference [68] with permission from ACS publications**

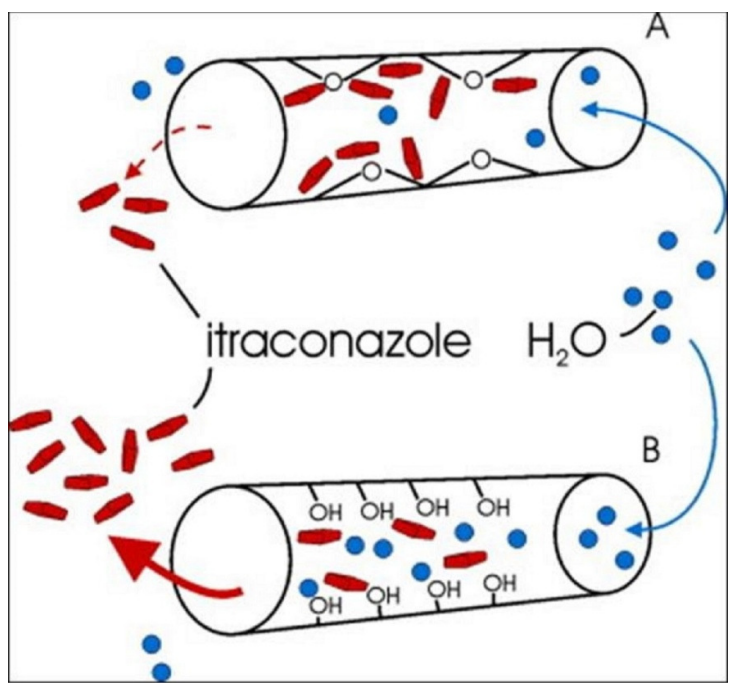

**Fig. 2. Effect of humidity on Ibuprofen loaded SBA-15 surfaces; reprinted from reference [69] with permission from Elsevier**

## 3.1 Use of non-ordered Mesoporous Silica in Convention Solid Dispersion as Adsorbent

Some scientists had used a combination of melt granulation and surface adsorption techniques to enhance the dissolution and tableting properties of poorly soluble drugs like Glibenclamide [80], Cefuroxime [81]. The solid dispersion of Cefuroxime used gelucire as melt dispersion carrier and Sylysia 350 as an adsorbent. The surface adsorbent, Sylysia 350 was used not only to impart good flow and compressibility to dispersion granules but also to improve the dissolution rate by increasing the effective surface area [81]. Another study was conducted using poorly soluble drug sulfathiazole and polyvinyl pyrrolidone as a carrier and Syloid 244FP as an adsorbent [82]. This method combined the advantages of conventional solid dispersion and surface solid dispersion by utilizing the adsorptive nature of porous silica particle and modifying the model drug into optimal molecular form [82]. This technology helps to prevent the recrystallization of the drug on storage by forming strong hydrogen bonds between the drug and polymer complex, which is adsorbed on the porous carrier with further stabilization effects of silanol group present on the pore surfaces. Due to the large surface area of this system, the dispersion is immediately dispersed as soon as it touches the dissolution media and prevents the crystal growth during dissolution and the higher dissolution rate was achieved [82].

## 3.2 Stability of Drug Loaded on Non-ordered Mesoporous Silica

Miura et al investigated the stability of poorly soluble drug K-832 using porous silica Sylysia 740 (2.5 nm diameter pores and Sylysia 350 (21 nm diameter pore) [83]. These formulations were stored in 60°C and 80% RH in open and closed condition. It was observed that formulation stored in 60°C and 80% RH closed condition was difficult to crystallize amorphous K-832 in the mesopores of the silica. However, the formulation stored in 60°C and 80% RH in open conditions showed a diffraction peak at low intensities in powder X-Ray diffraction analysis. DSC measurements revealed the difference in heat of fusion of K-832-Sylysia 740 formulation increases on storage; however, it is smaller than that of K-832-Sylysia 350 formulation. This result suggests that K-832 adsorbed on Sylysia 740, which has a smaller pore diameter, has a higher physical stability of amorphous K-832. Furthermore, molecular mobility was measured in relaxation time study using solid state $^{13}$C NMR, revealed that the molecular mobility of K-832 was lower for the 2.5 nm pore than for the 21 nm pores, thus making the crystallization rate of amorphous K-832 in the 2.5 nm pores much slower. The result is in agreement with the study conducted by Aso, et al. and Masuda, et al. [84,85]. They have correlated crystallization behavior of amorphous compounds with the difference in the molecular mobility of amorphous compounds, which were based on spin-lattice relaxation time measured by solid state $^{13}$C NMR spectroscopy based on Adams-Gibbs-Veogel equation.

## 4. METHOD OF DRUG INCORPORATION ON MESOPOROUS SILICA

Various methods are available for incorporation of drugs into mesoporous materials like solvent deposition methods [86-90], mechanical activation [70,91,92], vapor-phase mediated mass transfer [32, 93]. In spite of the preparation method, mesoporous silica systems has high surface area and high surface energy, and adsorption of the drug molecules on the porous material allows the system as a whole to progress to a lower free energy state i.e. Gibbs free energy is reduced and this type if systems are stable [18]. The drug exists as an amorphous material in these systems and crystallization of amorphous material will take place only if the thermodynamic state of the system is disturbed. In addition to thermodynamic factors, crystal growth and nucleation is hindered by spatial constraints, i.e. the pores are not able to incorporate enough molecules in order for them to reach a critical nucleation size [18,34,89].

The physical state of hydrophobic drug molecules in ordered mesoporous silica is influenced by the loading procedure. Mellaerts et al compared the drug incorporation methods using ordered mesoporous silica SBA-15 and model drugs ibuprofen and ITZ [94]. Three loading procedure were studied like adsorption from solution (solvent method), incipient wetness impregnation and heating of the drug and the SBA-15 physical mixture. ITZ was successfully dispersed in SBA-15 using adsorption from dichloromethane and incipient wetness impregnation method. ITZ molecules were molecularly deposited over micro- and mesopores at 20% drug loading. At the higher drug loading, an adsorbed layer is formed in which ITZ molecules interact in a way similar to glassy state. The solvent method favors the positioning of ITZ on mesopores wall and the incipient wetness impregnation methods favor positioning in the micropores. The sample prepared from solvent method shows fast release kinetics. Ibuprofen was successfully incorporated inside the micropores using melt method due to less viscosity in the molten state. Both ITZ and ibuprofen released fast from SBA - 15 when exposed to simulated gastric fluid and supersaturation solution are easily obtained. The loading of the ibuprofen and ITZ in ordered mesoporous silica by melt method depends on the drug molten viscosity, while the ability to form a homogeneous mixture of drug and silica prior to the melting step of the process depends on the density of the powders and method of blending.

Some scientist has used supercritical fluid techniques (SCF) to load the drug in mesoporous silica since solvent immersion method often leads to the pores of mesoporous materials not be fully utilized, and results in a low drug loading efficiency and an increase in dissolution rate [95]. Li-hong and coworkers loaded ibuprofen onto MCM-41 using solvent immersion method and supercritical carbon dioxide drug loading procedure [95]. It was found that the amount and the depth of ibuprofen entered the pores of the mesoporous silica by SCF technique were larger than those by the solution immersion method. A similar study conducted on Vitamin E [96] and results were in agreement with the previous study. The drug loading methods like solvent immersion, supercritical carbon dioxide, and liquid carbon dioxide, were able to convert crystalline drug in the amorphous state [83,97-99]. The spontaneous transition of crystalline to the amorphous state of drug occurred in physical mixes of drug and silica via a vapor phase-mediated pathway for drugs with a relatively low vapor pressure [93]. This unique phase transformation phenomena are a simple and effective way to improve the aqueous solubility and bioavailability of poor water soluble drugs.

## 5. EFFECT OF SURFACE FUNCTIONALIZATION

The drug loading is governed by interfacial interactions between the drug molecules and the surface of the mesoporous silica. The surface chemistry of the mesoporous silica determines the two common mechanisms like physical and chemical adsorption. Generally, if one seeks enhanced drug dissolution, physical adsorption like hydrogen bonding, electrostatic and hydrophobic interactions are preferred. The unmodified mesoporous silica is covered by –OH groups, hence hydrogen bonding are the most common interaction [38,39]. Song et al. [100] studied the functionalized mesoporous silica SBA-15 with the amine group through post-synthesis and one pot synthesis on model drug ibuprofen and bovine serum albumin. Fig 3 represent the TEM images of surface functionalized mesoporous silica. It was found that the release rate of ibuprofen functionalized by post synthesis can be effectively controlled as compared to that from pure SBA-15 and SBA-15 functionalized by one-pot synthesis due to ionic interaction between the carboxyl group in ibuprofen and an amine group on the surface of SBA-15. However, the release rate of bovine serum albumin functionalized by one-pot synthesis is more favorable due to the balance of electrostatic interaction and hydrophilic interaction between the bovine serum albumin and the functionalized SBA-15 matrix. Similar results were observed for mitoxantrone, the release rate is strongly dependent on the pH of the release medium and the type of surface functional group [101]. Conversely, Vallet-Regi and co-workers found that ibuprofen loading was higher in non-functionalized MCM-41 silica compared to amine-modified silica [102]. A similar trend was observed for erythromycin [103].

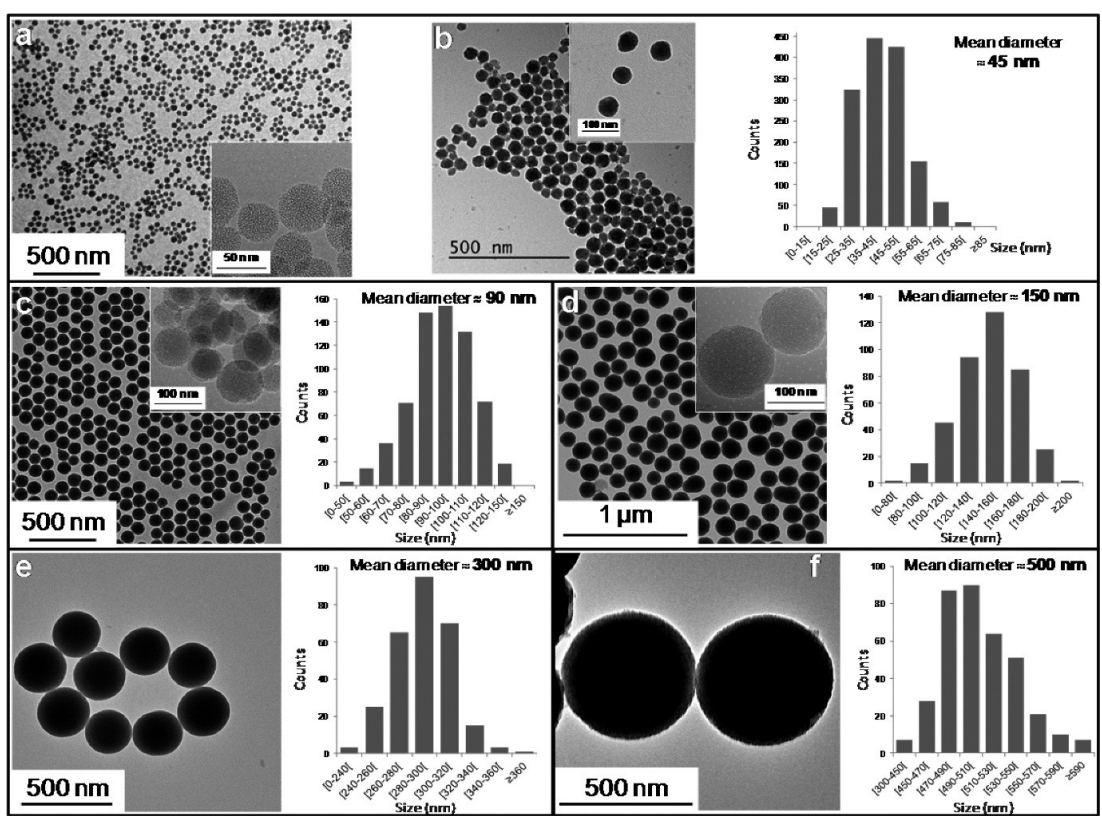

**Fig. 3. TEM images of the different surface functionalized mesoporous silica, with their corresponding size distribution: (a and b) MSN45; (c) MSN90; (d) conventional MSN150; (e) MSN300; and (f) MSN500. The insets (in a, c, and d) are HRTEM images, reprinted from reference [104] with permission from Elsevier**

# 6. EFFECT OF PORE SIZE AND PARTICLE SIZE

The pore size can determine how large molecules can be loaded to the carrier molecule, and it is very crucial for drug loading because of the mesopores in mesoporous silica act as sieves. The pore should be accessible for drug molecule and in general, the pore diameter/drug molecule size should be >1. For high drug loading in the mesoporous material, the ratio of pore diameter/drug molecule size should be greater than 3. Shen et al. [90] studied the effect of pore size and particle size of ordered mesoporous silica material on model drug ibuprofen. Mesoporous silica material (MCM-41 and SBA-15) with varied pore and particle size was co-spray dried at high drug loading with ibuprofen. The solid dispersion obtained has a significant enhancement in the dissolution rate, which was affected by physical state and particle size of ibuprofen in mesoporous structure. Amorphous state of ibuprofen was obtained when ibuprofen was co-spray dried with MCM-41 and SBA-15 with a pore size smaller than 10 nm due to nano space confinement. In contrast, when ibuprofen is co-spray dried with SBA-15 with a pore size of above 20 nm nanocrystal were obtained and it had a slower dissolution than SBA-15 and MCM-41. The particle size of mesoporous silica showed less than a pronounced effect on the dissolution of ibuprofen. Vallet-Regi [40] use the hydrothermal effect to make mesoporous silica with different pore size by use of different surfactant. The swelling effect of the surfactant template which ultimately leads to increase in pore size of SBA-15 has increased from 8.4 mn up to 11.4 nm due to the increase in pressure. Control of the pore size and morphology of porous silica drug carriers have been found to affect the drug dissolution profile [105-107]. The release kinetics of confined drugs depends on pore diameter and the particle size/morphology, drug release may occur in the order of minutes, hours or even days.

Horcajada et al. [108] also studied the effect of pore size of mesoporous silica MCM-41 on the drug loading on ibuprofen. MCM-41 was prepared with different pore sizes by using surfactant molecules with different lengths of the alkyl chain. Ibuprofen was adsorbed on mesopores of MCM-41 with varying pore size, more ibuprofen molecules were loaded into the carrier with the pore size of 3.6 nm (19% wt drug loading) than with a pore size of 2.5 nm (11% wt drug loading). Ibuprofen loaded on a carrier with 3.6 nm pore size, a monolayer of drug molecules fully occupied the pore surface and the space in the middle of the pore was available to allow the drug molecules to freely diffuse as seen in Fig. 4. As the pore size decreased to 2.5 nm, due to geometrical consideration (steric hindrance) the close packing of the drug molecule along the pore wall was avoided. Zhang et al. [109] studied the another poorly soluble drug telmisartan on the spherical mesoporous silica, increasing pore size from 3.6 nm to 12.9 nm have increased the drug loading to 59.7% in 12.9 nm pores. Kinnari et al. [87] studied non-ordered mesoporous silica (Syloid AL-1 and 244) microparticle as a carrier for a hydrophobic drug ITZ. The drug loading was performed by the immersion method, in which the particles were immersed into concentration ITZ solution in dichloromethane (high and low drug concentration). The two syloid showed similar drug loading capacities when low drug concentration was used despite their significant differences in the pore size, surface area, and pore volume. At high loading concentrations of ITZ, the loading capacities for syloid AL-1 and 244 was increased to 25.1 and 32.8 w-% respectively. The higher drug loads observed for the mesoporous silica particles might be attributed to different pore geometry. The surface structure of the mesoporous silica contains a large amount of silanol groups that are favorable for interactions with ITZ through hydrogen bonds. ITZ was present in an amorphous form which resulted in a significant improvement in dissolution. The amorphous form of ITZ was confirmed by DSC, XPRD measurements, SEM. Further, the FTIR investigations revealed that loading process did not change the chemical structure or morphology of the particle surface. The release of ITZ at pH 1.2 from Syloid AL-1 microparticle was found to be faster than Syloid 244 particles.

# 7. EFFECT OF SOLVENT

Drug loading is influenced by the polarity of the solvent. Dimethyl sulfoxide (DMSO) is very polar solvent and it can form competitive adsorption with the drug molecules causing a low degree of drug loading. Charnay et al. [110] systematically studied the effect of solvent (DMSO, dimethylformamide (DMF), dimethylacetamide (DMA), ethanol and hexane) on drug loading of poorly soluble drugs. It was revealed that ibuprofen was not loaded into the MCM-41 mesoporous material when DMA is used as a solvent, this is the result of the extreme polarity of the DMA solvent. However, in hexane, the loading capacity was up to 37% wt. The solvent plays an important role on the drug loading with mesoporous material based on TiO$_2$ and Al$_2$O$_3$.

## 8. DRUG RELEASE MECHANISM

To design the optimal carrier system one must have a thorough understanding of drug release kinetics. The release profile of poorly soluble drugs can be fitted with Higuchi equation [111] (Eq. (1)).

$$Q = Kt^{1/2} \tag{1}$$

Where Q is the cumulative amount of drug release at time t, and K is the Higuchi constant.

The release of drug molecules from the insoluble carrier is described by Higuchi equation. Since the mesoporous materials are manufactured using inorganic solvents and compound, they are not soluble in the aqueous solution under the biological condition and hence, as suggested by Higuchi equation the drug release from inorganic material is the diffusion controlled process [111]. Hence Higuchi equation can be applied to the carrier of the mesoporous system to explain the drug release kinetics [108,112,113]. Zhao et al [112] showed a pure Higuchi type of diffusion driven drug release with hollow magnetic mesoporous silica carrier on model drug ibuprofen. The results suggest the presence hydrogen bonding between the drug and mesoporous material. Hydrogen bonding is the weak intermolecular forces and hence it does not control the dissolution of the drug molecules from the mesoporous material. However, Anderson et al [113] observed two-step release phenomenon with ordered mesoporous silica carrier according to the fitting of the release data with Higuchi equation due to the degradation of mesoporous carrier MCM-41 and SBA-3. Nevertheless, the deviation from Higuchi equation is less pronounced for the more stable mesoporous carrier. Similar two step release mechanism was observed for thermally stabilized PSi (TCPSi, thermally carbonized PSi and TOPSi, thermally oxidized PSi) and non-ordered mesoporous silica (Syloid AL-1 and 244) (Fig. 5) [87]. Kinnari et al showed that in the first stage, the physically entrapped drug molecule is released faster and in the second stage slower release rate is observed due to chemical bonding between the drug and the surface of the mesoporous materials which sufficiently agreed with the Higuchi model (Fig. 5). Some scientist showed that the drug molecule located on or near the surface of the porous material can be released quickly due to instant dissolution and release in medium, whereas the drug molecules packed deeper inside the pore network are released slower [72,114].

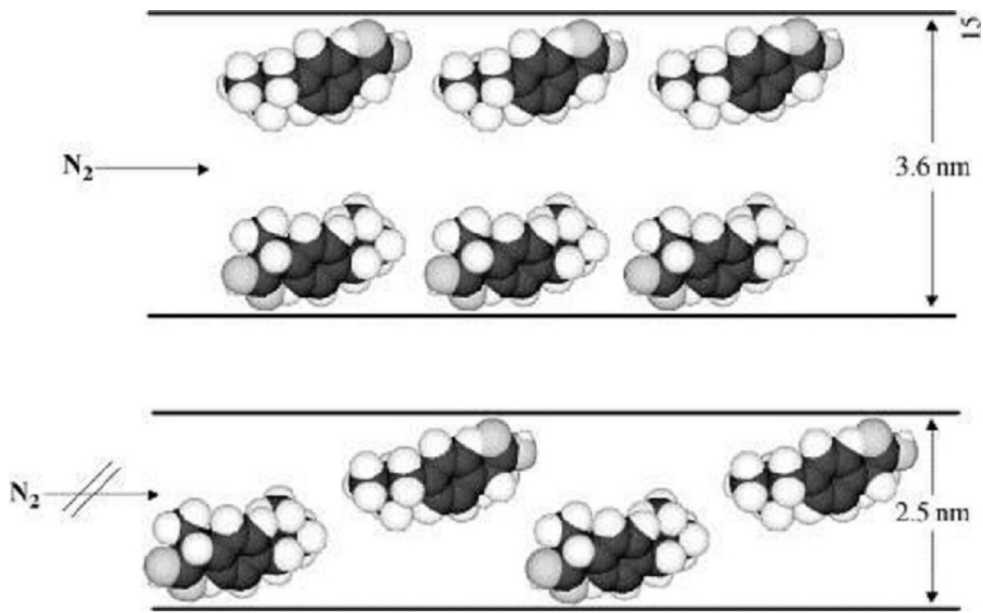

**Fig. 4. Ibuprofen molecules inside the pore channels of different pore sized MCM-4; reprinted from reference [108] with permission from Elsevier**

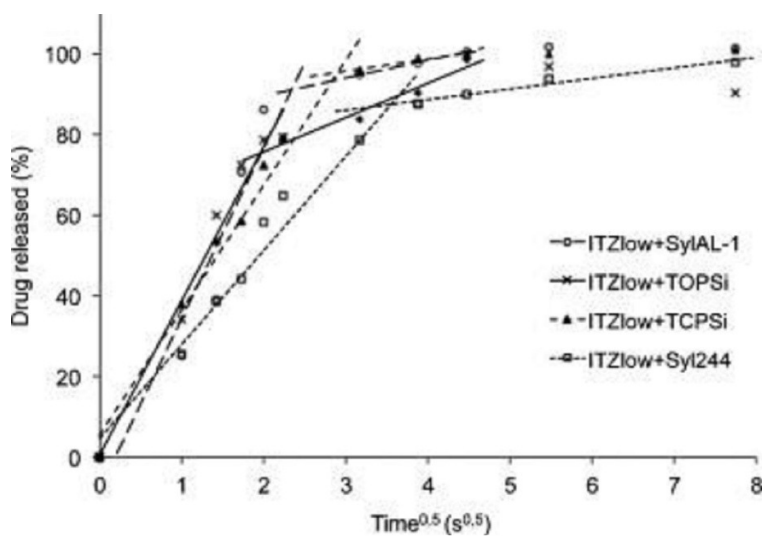

**Fig. 5. Release of itraconazole at pH 1.2 fitted with Higuchi model; reprinted from reference [87] with permission from Elsevier**

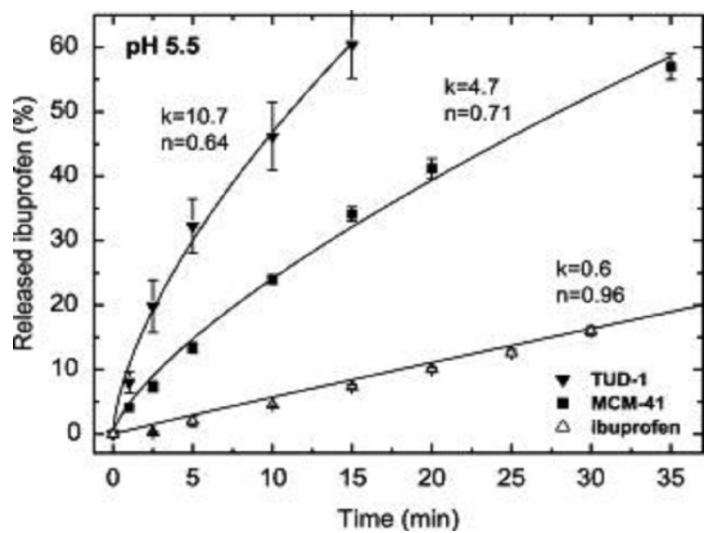

**Fig. 6. Release profile of ibuprofen (initial 60%) at pH 5.5 fitted with Korsmeyer–Peppas model; Reprinted from reference [116] with permission from Elsevier**

The drug release kinetics from the mesoporous carrier can be described in a more comprehensive way by Korsmeyer –Peppas equation (Eq. ( 2)).

$$F = \left(\frac{Mt}{M}\right) = K_m t^n \qquad (2)$$

Where, F is the fraction of the drug release at time t, M is the total amount of loaded drug in the mesoporous carrier; $K_m$ is the kinetic constant and n is the release constant which describes the drug release mechanism [115].

If n= ½ in Eq. (2), this model is reduced to Higuchi model describes diffusion controlled release. The drug release characteristics of ibuprofen loaded TUD-1 and MCM-41 the release profile is fitted with the Korsmeyer-Peppas model [116] is shown in Fig. 6. The kinetic constant of K=10.7 was observed for the TUD-1 carrier, which is higher than MCM-41 (k=4.7) demonstrated the unrestricted release of the drug from release medium due to high accessibility and stability of the TUD-1-1 mesoporous network. The modeling of the dissolution curve of the pure crystalline form of ibuprofen shows the much slower release of the drug (k=0.6) compared to the mesoporous carrier. The release constant n revealed that the ibuprofen release mechanism of the TUD-1(n=0.64) material was more diffusion based than the MCM-41 material (n=0.71). The highly accessible nano reservoir of the TUD-1 material provided a relatively unrestricted release of ibuprofen, whereas the long and narrow mesopores pathway of MCM-41 sterically hindered the free diffusion of ibuprofen from the mesopores. However, Korsmeyer-Peppas model was found to describe only the first 60% of the drug release [116,117].

The mesoporous silica carrier also affects the permeation properties of the drug molecules. The mesoporous silicon nanoparticle loaded with telmisartan significantly enhance telmisartan permeability and reduce the drug efflux [118]. The cellular uptake of mesoporous silica is highly time, concentration and size dependent.

# 9. In vivo STUDIES

The poor correlation between in vitro and in vivo results emphasizes the importance of animal tests when developing formulations with poorly soluble drugs. The poorly soluble drugs are successfully loaded into mesoporous silica as an amorphous or disordered form. The enhanced drug dissolution is observed when using mesoporous silica as a carrier. Hence we expect it to have enhanced bioavailability [40,48,97]. The in vivo performance of ordered mesoporous silica as a carrier for ITZ was evaluated in rabbits and dogs [40]. After oral administration of ordered mesoporous silica loaded with ITZ to dogs, the area under the curve 0-8 h ($AUC_{0-8}$) was $681 \pm 566$ nM while no systemic ITZ was detected when pure crystalline ITZ was administered. In rabbits, after oral administration of ordered mesoporous silica loaded with ITZ, $AUC_{0-24}$ was increased to $1069 \pm 278$ nM which was double of that of crystalline ITZ. When SBA-15 is used as a mesoporous carrier, tmax was decreased to $4.2 \pm 1.8$ h. The pharmacokinetic parameters obtained with SBA-15 were comparable to commercially available Sporanox®. Zhang et al. [109], studied the release and bioavailability of telmisartan loaded in the nano and microparticles of mesoporous silica, Fig. 7. Both nano and micro particle were compressed into tablets and the drug release rate was studied (MSN -nanoparticle tablet and,

MSM-microparticle tablet). Fig. 7 shows the in vivo mean plasma concentration of telmisartan following an oral dose of a Micardis tablet and telmisartan loaded in the nano and microparticles of mesoporous silica in beagle dogs. The release rate is significantly improved in comparison with commercially available Micardis. The results were confirmed with subsequent in vivo studies in dogs. The AUC 0-72 h was approximately 1.29 times greater when MSM was administered and the mean value of Cmax for MSN was 1,24 and 1,13- fold greater than that of the commercially tablet and MSM. These results have shown that use of mesoporous material increases the bioavailability of the poorly water soluble BCS class II and IV drugs. The drug loaded in mesoporous silica create a supersaturated drug solution in vivo which has a tendency to crystallize out [41] and hence the addition of excipients which reduces precipitation will be useful. Some scientists have incorporated polymeric materials in amorphous formulations to inhibit the precipitation and in vivo studied were conducted in rats [48]. In this study, combined use of SBA-15 and precipitation inhibitor like HPMC and HPMCAS was used to improve oral absorption of ITZ. In vitro studies indicated that HPMC and HPMCAS inhibited the crystallization up to 4 hours. Surprisingly, in vivo studies carried out showed lower AUC of a formulation containing HPMCAS than formulation without HPMCAS. The formulation with HPMC showed the higher AUC than the formulation without HPMC. However, Vandecruys et al showed that precipitation kinetics cannot be effectively inhibited by excipients [119]. Further, the selection of excipients mediated precipitation inhibition is dealt on the empirical basis [43,119] and structure activity relationship is not yet established. It is well understood that degree of supersaturation and rate at which supersaturation is created affects the rate and mechanism by which precipitation occurs [120]. This phenomenon was studied on model drug fenofibrate in which ordered mesoporous silica SBA-15 and MCM-41 of varying pore diameters were loaded with fenofibrate. The in vitro release rate analysis showed that increase in release rate as we increase the pore size. However, the In vivo experiments conducted in rats showed the bioavailability of fenofibrate increases as we decrease the pore size [121]. More understanding is needed to gain a better understanding.

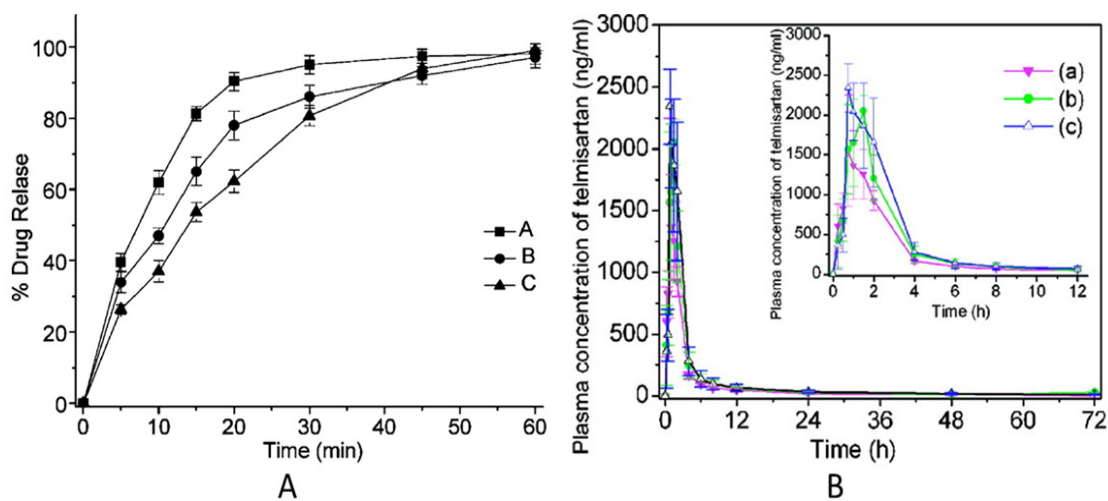

**Fig. 7A. Release profile of telmisartan (TEL) from A) TEL-loaded MSN tablet, (B) TEL-loaded MSM tablet and (C) TEL commercial tablet in Phosphate buffer solution (pH6.8); Reprinted from reference [118] with permission from ACS publication; Fig. 7B: *In vivo* mean plasma concentration versus time curves of TEL following an oral dose of (a) a Micardis tablet, (b) a TEL-MSM tablet and (c) a TEL-MSN tablet (equivalent to 40 mg of TEL) in beagle dogs. Each data point represents the mean ± SD (n = 6); Reprinted from reference [118] with permission from ACS publication**

## 10. CONCLUSION

Mesoporous silica is a promising candidate to improve the release profile of poorly soluble drugs. Since the drug loading is taking place by physical adsorption and subsequent pore filling a wide variety of drugs can be used and dosage forms can be prepared using a downstream process like compression or capsule filling. Although several *in vitro* studies showing the enhanced dissolution behavior by use of mesoporous silica are carried out, but there are relatively few in vivo studies on this topic. Hence, all the features of the mesoporous carrier have not been thoroughly explored. These mesoporous carriers not only improve the dissolution but also permeation of the drug across the gastrointestinal membrane. The data in the literature have demonstrated the scientific and commercial promises of using mesoporous silica as an amorphous delivery system. Amorphization of the drug loaded on the mesoporous materials occurs through the limitation of space i.e. if the pore size of the mesoporous material is smaller than the critical nucleation size which will eventually lead to stabilization of the amorphous drugs by decreasing mobility of the drug molecule by incorporating them in the pore of the mesoporous material. The system is physically stable due to a decrease in Gibbs free energy. However, more work is needed to understand the detailed mechanism of interactions of drugs in mesoporous silica. There is very little information available about the molecular arrangements and vibrational spectroscopy, which is used to gain information on molecular interactions such as hydrogen bonding and intermolecular forces and bonding between the drug molecule and the mesoporous material. Commercialization of mesoporous silica formulation is possible if the mechanism of interaction is studied in dept. Silica based material like aerosil have already been used as pharmaceutical excipients for several years and hence the obstacle for commercialization of the technology is minimal.

## CONSENT

It is not applicable.

## ETHICAL APPROVAL

It is not applicable.

## COMPETING INTERESTS

Authors have declared that no competing interests exist.